\documentclass[traditabstract]{aa}

\usepackage{txfonts}
\usepackage{graphicx}
\usepackage{natbib}

\newcommand{\mina}{^{\,\prime}}
\newcommand{\seca}{^{\,\prime\prime}}

\newcommand{\kms}{$\rm km\,s^{-1}$}
\newcommand{\am}{$^{\prime}$}
\newcommand{\as}{$^{\prime \prime}$}

\newcommand{\hi}{\ion{H}{I}}
\newcommand{\hii}{\ion{H}{II}}

\newcommand{\G}{GSH\,91.5+2$-$114}

\newcommand{\IRdos}{IRAS\,21147+5016}
\newcommand{\IRtres}{IRAS\,21093+4959}
\newcommand{\IRcuatro}{IRAS\,21041+5052}

\newcommand{\mm}{G91.56+0.97}
\newcommand{\nn}{G92.24+1.57}

\begin{document}

\title{GSH 91.5+2$-$114\\ A large \hi\, shell in the outer part of the Galaxy} 

\author{S. Cichowolski\inst{1,2} \and S. Pineault\inst{2,3}}

\institute{Instituto de Astronom\'{\i}a y F\'{\i}sica del Espacio (IAFE),
CC 67, Suc. 28, 1428 Buenos Aires, Argentina
\and
D\'epartement de physique, de g\'enie physique et
d'optique, Universit\'e Laval, Qu\'ebec, G1V 0A6 Canada, and Centre de recherche
en astrophysique du Qu\'ebec (CRAQ)
\and
Instituto Argentino de Radioastronom\'{\i}a (IAR),
CC 5, 1894, Villa Elisa, Argentina}

\date{Received date / Accepted date}

\abstract
{GSH~$91.5+2-114$ is a large \hi\ shell located in the outer Galaxy at a kinematic
distance of  about 15 kpc.  It was first identified in the Canadian Galactic Plane Survey
(CGPS) by Pineault et al. (2002) as being possibly associated with
objects possessing infrared colors which indicates strong stellar winds.
The \hi\ shell has no obvious continuum counterpart in the CGPS radio images at
408 and 1420 MHz or in the IRAS images.  We found no evidence for 
 early-type massive stars,  most likely as a result of the large extinction that is expected for
this large distance.  An analysis of the energetics and of the main
physical parameters of the \hi\ shell shows that this shell is  likely the result of
the combined action of the stellar winds and supernova explosions of many stars.
We investigate whether a
number of slightly extended regions characterized by a thermal radio continuum and located
near the periphery of the \hi\ shell could be the result of star formation
triggered by the expanding shell.}

\titlerunning{A large \hi\ shell in the outer part of the Galaxy}
\authorrunning{S. Cichowolski \& S. Pineault}

\maketitle

\section{Introduction}

A massive star possessing a strong stellar wind (SW) can inject as
much energy into the interstellar medium (ISM) during its lifetime as it
does during the final supernova explosion.  
If the stellar spatial velocity with
respect to the ambient ISM is not too large (i.e. $<$ 30 \kms), the SW is
expected to evacuate a large cavity around the star (called a stellar
bubble) surrounded by a ring or shell of enhanced density.  These
structures have been observationally detected as optical and infrared
(IR) nebulae around Wolf-Rayet (WR) and Of stars \citep{chu83, loz92}.

The situation with respect to radio observations, reviewed  by
\citet{cap03}, has evolved considerably with the advent of large
Galactic neutral hydrogen (\hi) surveys, namely the Canadian Galactic Plane Survey
\citep[CGPS;][]{tay03}, the Southern GPS \citep[SGPS;][]{mcc05,hav06}
 and the VLA \citep[VGPS;][]{sti06}. 
These surveys
at $\sim 1$ arcmin resolution have made it possible to identify and
study the structure and dynamics of many neutral \hi\ shells
\citep{mcc00,uya02,sti01,sti04}.  Interestingly enough, the large majority of
shells detected by their neutral hydrogen emission have low inferred
expansion velocities, typically less than or on the order of 10 \kms\, 
\citep{cap03}.

In parallel to these developments on the observational scene, recent
theoretical studies, building up on the initial work of \citet{wea77} and others, have
considerably increased our understanding of the interaction of stellar
winds with their surrounding ISM.  The effects of the different
evolutionary phases and of a large peculiar motion of the star have
been modeled in detail in a number of new studies \citep[e.g.,][and references therein]{art06, bri95, van88, wil96}. Concerning
the often observed low expansion velocities of $\sim 10 \, \rm
km\,s^{-1}$, \citet{caz05} have shown that velocity
dispersion within the shell and the role of the local ISM background
may significantly affect the appearance of an expanding \hi\ shell in
velocity space and thus its inferred parameters (in particular, 
mass and expansion velocity).

In the case of moderate-size shells, a 
puzzling aspect is the apparent lack of a radio continuum
counterpart or of a candidate progenitor star \citep{nor00, sti01, sti04, cic04}.  
This suggests that either one or several basic ingredients
are missing in the predictions of the theory and/or that some detected
\hi\ shells are not real.

Increasing the sample of well studied \hi\ shells is a first natural
step in elucidating some of the current puzzles.  In an attempt at
diversifying the sample of known \hi\ shells (consisting mostly of
shells around objects known for their optically interesting features),
\citet{pin02} initiated a project aimed at
discovering new SW source candidates by using first IRAS colors to
extract potential candidates and then the CGPS database to look for a
morphology indicative of a SW, i.e.~shells, rings, bubbles, cavities,
or voids. An obvious advantage of this procedure is that potential
candidates suffer much less from the selection effects associated with
optically chosen targets, for example, WR or Of stars, the
distribution of which is severely biased by absorption. A positional
coincidence between one or more candidates and a shell-like morphological
structure is nevertheless not a proof of a physical association, and 
a detailed analysis is required before any firm conclusion can be
drawn.

In this paper, we focus our attention on a very large (nearly 1\fdg5
diameter) and symmetrical \hi\ shell centered at ({\it l, b}) = 
(91\fdg5, +2\fdg0) and observed at a velocity\footnote{All velocities
are with respect to the local standard of rest (lsr)} $v_{\rm lsr} \approx
-114$ \kms,  which was identified by \citet{pin02}.
A flat rotation curve model for the Galaxy gives
for the shell a distance $d = 15 \rm\, kpc$, a galactocentric radius
$R_g = 17 \rm\, kpc$,  a distance $z \sim$ 525 pc from the Galactic plane, and a diameter $D = 400 \rm\, pc$, placing it in
the outermost part of the Galaxy.

This shell thus offers the opportunity to explore the environment
of remote regions of the Galaxy where many physical
parameters such as metallicity, density, smaller or negligible 
perturbations from spiral arms, greatly differ from those in nearby 
regions of the Galaxy.  Despite this remoteness, there is evidence
that star formation in the outer Galaxy may be common, as shown by the
discovery of a considerable number of embedded star clusters in molecular
clouds up to galactocentric radii $R_g \approx 17 \rm\, kpc$
\citep{san00,sne02}.  A particularly interesting case is the suggestion
by \citet{kob08} that star formation in Digel's Cloud 2 ($R_g \approx 19 
\rm\, kpc$) could have been triggered by the huge supernova remnant 
GSH~138-01-94 previously discovered by \citet{sti01}.  These huge
shells imply that {\em massive} stars form in the outer Galaxy, emphasizing
the importance of studying these objects.  At these large distances, optical
obscuration is very severe, so that one has to resort to
radio or infrared observations.

The plan of the paper is as follows.  In Section \ref{obs} we describe
the observational data used, in Section \ref{results} we briefly
review the initial data used by \citet{pin02}
and present detailed neutral hydrogen (\hi) and
continuum images of the object.  The results are analyzed and discussed
in Section \ref{discuss}.  Section \ref{concl} is a summary of the main 
conclusions.

\section{Observational data}\label{obs}

Radio continuum data at 408 and 1420 MHz and 21-cm spectral
line data were obtained at DRAO as part of the CGPS survey \citep{tay03}.
A detailed description of the data processing routines can be found in \citet{wil99}.
At the position of \G, the continuum images have a resolution of $1\farcm08 \times 0\farcm82$
and $3\farcm65 \times 2\farcm82$, and a measured noise of 0.082 K and
0.75 K,  at 1420 and 408 MHz respectively.
Parameters relevant to the \hi\ data are given in Table \ref{hidata}.
High-resolution-processed (HIRES) \citep{fow94}
{\em IRAS} images produced at the Infrared Processing and Analysis
Center (IPAC)\footnote{The Infrared Processing and Analysis Center
(IPAC) is funded by NASA as part of the {\em Infrared Astronomical
Satellite (IRAS)} extended mission under contract to the Jet
Propulsion Laboratory (JPL)} were also used. The images used are the result of 20
iterations of the algorithm, giving an approximate resolution ranging from
about $0\farcm5$ to $2\mina$.  At the position of \G, the 60 micron image has
an approximate resolution of $1\mina$.

\begin{table}
\caption{Observational parameters for the \hi\ data} 
\label{hidata}
\centering
\begin{tabular}{lc}
\hline\hline
Parameter &  Value\\
\hline
Synthesized beam  &  1\farcm26 $\times$ 0\farcm98 \\
Observed rms noise (single channel)(K) & 1.6  \\
Bandwidth (MHz) &  1.0 \\
Channel separation (\kms) &  0.824 \\
Velocity resolution (\kms) & 1.32 \\
Velocity coverage (\kms) & 224\\
\hline
\end{tabular}
\end{table}

\section{Results}\label{results}

\subsection{The initial data}\label{initdata}

\begin{figure}
\resizebox{\hsize}{!}{\includegraphics{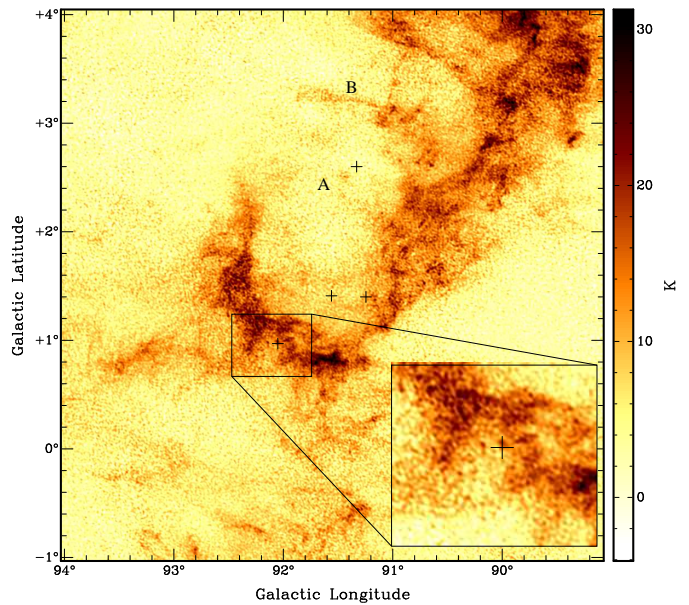}}
\caption{\hi\ emission distribution averaged between $-110.3$ and $-116.9$
\kms. The plus symbols indicate the position of the four DWCL source candidates, 
IRAS\,21147+5016, IRAS\,21106+5013, \IRtres, and \IRcuatro. 
The inset is a close-up view of the region near \IRdos.}
\label{average}
\end{figure}

Figure \ref{average} shows 
an \hi\ image averaged between $v = -110.3 $ and $-116.9$ \kms,
showing the large \hi\ shell discovered by \citet{pin02} with the four IRAS sources which they suggested
might be physically associated.

The interior of the shell seems to have been entirely cleared of
neutral hydrogen.  Though the shell is generally quite well defined, 
its northern part is essentially absent, or at least it does not
show a clear outline. Indeed, the general morphology
suggests that the shell is open to the north in a direction 
away from the Galactic plane.
Indicated on Fig.~\ref{average} are two filaments (denoted
A and B) that could be related to the shell, forming an
incomplete northern border.
A closer inspection of the figure also shows
that the \hi\ shell is far from homogeneous. Moreover, there are many
quite well defined cavities visible that are projected onto the \hi\ ring.

The position of the four IRAS
sources is indicated by plus symbols in Fig.~\ref{average}. 
Two of the sources are seen projected well inside the shell structure,
while \IRtres\ is located between  filaments A and B.
\IRdos\ is seen projected onto a smaller \hi\ cavity centered at ({\it
l, b}) = (92\fdg1, 0\fdg9) and located on the larger shell structure.
The inset in the bottom right corner of Fig.~\ref{average} shows this region
in more detail.
Table \ref{colores} shows the IR colors of these four sources.

\begin{table*}
\caption{Parameters of the DWCL source candidates}
\label{colores}
\centering
\begin{tabular}{ccccccc}
\hline\hline
IRAS source  & $l$  & $b$  & $C_{12}$ &  $C_{23}$  & $F_{\rm ir}$(Jy) & $L_{\rm ir}(10^3 L_{\odot})$\\ 
\hline
21147+5016 & 92\fdg05  & 0\fdg97 &  $-0.34$ & 0.21  & $9.5 \pm 0.2$ & $3.0 \pm 1.2$\\
21106+5013 & 91\fdg56 & 1\fdg41 & $-0.32$ & 0.02  & $11.2 \pm 0.2$ & $4.0 \pm 1.6$\\
21093+4959 & 91\fdg25 & 1\fdg39 &  $-0.34$ & 0.16 & $9.2 \pm 0.2$ & $3.3 \pm 1.3$\\
21041+5052 & 91\fdg34 & 2\fdg60  & $-0.32$ & $-0.33$ & $27.1 \pm 0.2$ & $9.6 \pm 3.8$ \\
\hline
\end{tabular}
\begin{list}{}{}
\item  DWCL refers to a dusty late-type Wolf-Rayet star.
 \item $C_{ij} = \log (F_j/F_i)$, where $i$ and $j$ run from
1 to 4 and correspond to 12, 25, 60 and 100 $\mu$m,
respectively and $F_i$ is the flux density (Jy) in band $i$. 
\item  $L_{\rm ir} / L_{\odot} = 1.58\, F_{\rm ir}\, d_{\rm kpc}^2$ ,
where the integrated flux in Jy is $F_{\rm ir} =
1.3 (F_{12} + F_{25}) + 0.7 (F_{25} + F_{60}) + 0.2 (F_{60} +
F_{100})$ \citep{cha95}.  Subscripts are wavelengths in
microns . Luminosity values are for $d = 15$ kpc.
\end{list}
\end{table*}

A comparison of these colors with the values given by \citet{coh95}
in his Table~4 
\footnote{Cohen's (1995) definitions are of the form
$[i] - [j] = k_{ij} + 2.5 \log (F_j/F_i)$ where the constants
$k_{ij}$ are 1.56 and 1.88 for the colors $[12] - [25]$ and $[25] - [60]$,
respectively.}
allows us to conclude that the IRAS sources are probably dust shells related to
WC8-9 stars (dubbed {\it  dusty late-type WR}  or DWCL by Cohen). In their
preliminary analysis of the then-incomplete CGPS, \citet{pin02}
had concluded that the surface distribution of the stellar
wind candidates chosen on the basis of their IR colors was on the average
of 1 source per 10 square degree area.  Note that there are no other sources
with the same properties in the entire field shown in Fig.~\ref{average}.
The position of the four candidate sources inside or near the boundary 
of the \hi\ shell prompted a more detailed analysis.

\subsection{Single-channel \hi\ 21-cm data}\label{singlehi}

Figure \ref{hiset} shows the \hi\ distribution in the LSR velocity
range from $-107.0$ to $-121.0$ \kms. The original images were
smoothed to a 3-arcmin resolution to increase the signal-to-noise
ratio.  At $v = -109.5$ \kms\, the shell, which we shall now refer to as 
GSH\,91.5+2$-$114, is clearly observed and it is well defined down to $v =
-118.5$ \kms. This structure is centered at ({\it l, b}) = (91\fdg5,
+2$^{\circ}$) and has an angular size of  about  1\fdg5.

A limited region of bright centrally-located emission at --107.0 \kms\ could
form part of the so-called receding cap, implying an expansion velocity of
some 7 \kms.
In an ideal situation, one would then expect to find 
an approaching cap at a velocity of about --121 \kms. No obvious excess 
\hi\ emission is seen near this velocity.
The lack of confusing emission at this velocity suggests that
either expansion on the near side took place in an extremely low-density medium,
making its detection below the sensitivity of the telescope,
or that the shell was very incomplete and nearly absent on the near side.

\begin{figure*}
\centering
\includegraphics[width=17cm]{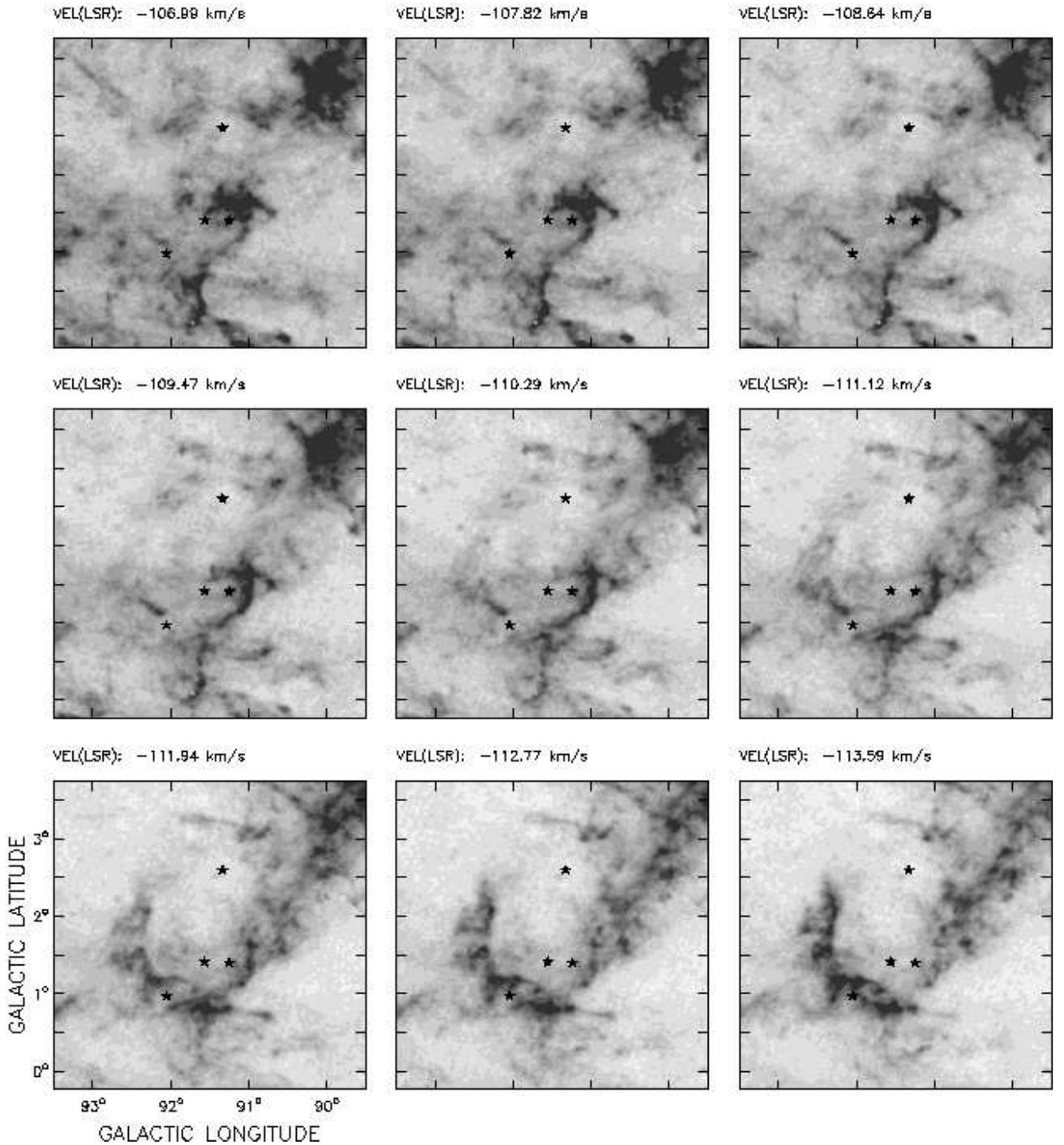}
\caption{CGPS gray-scale images showing the \hi\ distribution in the LSR
velocity range from $-107.0$ to $-121.0$ \kms. Velocity resolution is 1.3
\kms, and spatial resolution is 3\am. The LSR central velocity of each
image is indicated in the top left corner. The star symbols indicate the position of the four DWCL source candidates. For all images, darker
shading indicates higher brightness. }
\label{hiset}
\end{figure*}

\addtocounter{figure}{-1}
\begin{figure*}
\centering
\includegraphics[width=17cm]{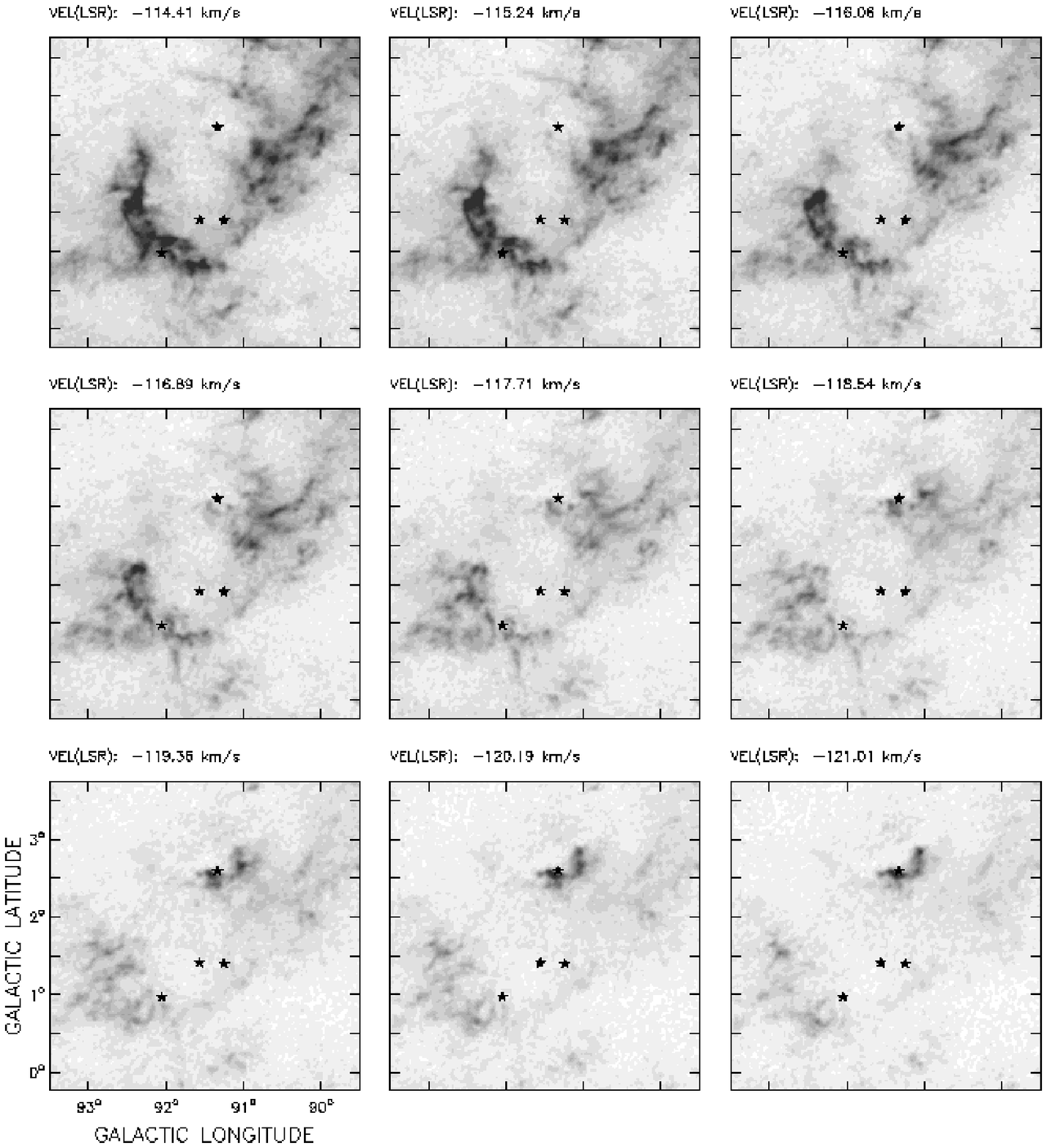}
\caption{{\it cont.}}
\label{hiset}
\end{figure*}

\subsection{Radio continuum and infrared data}\label{radir}

\begin{figure*}
\centering
\includegraphics[width=17cm]{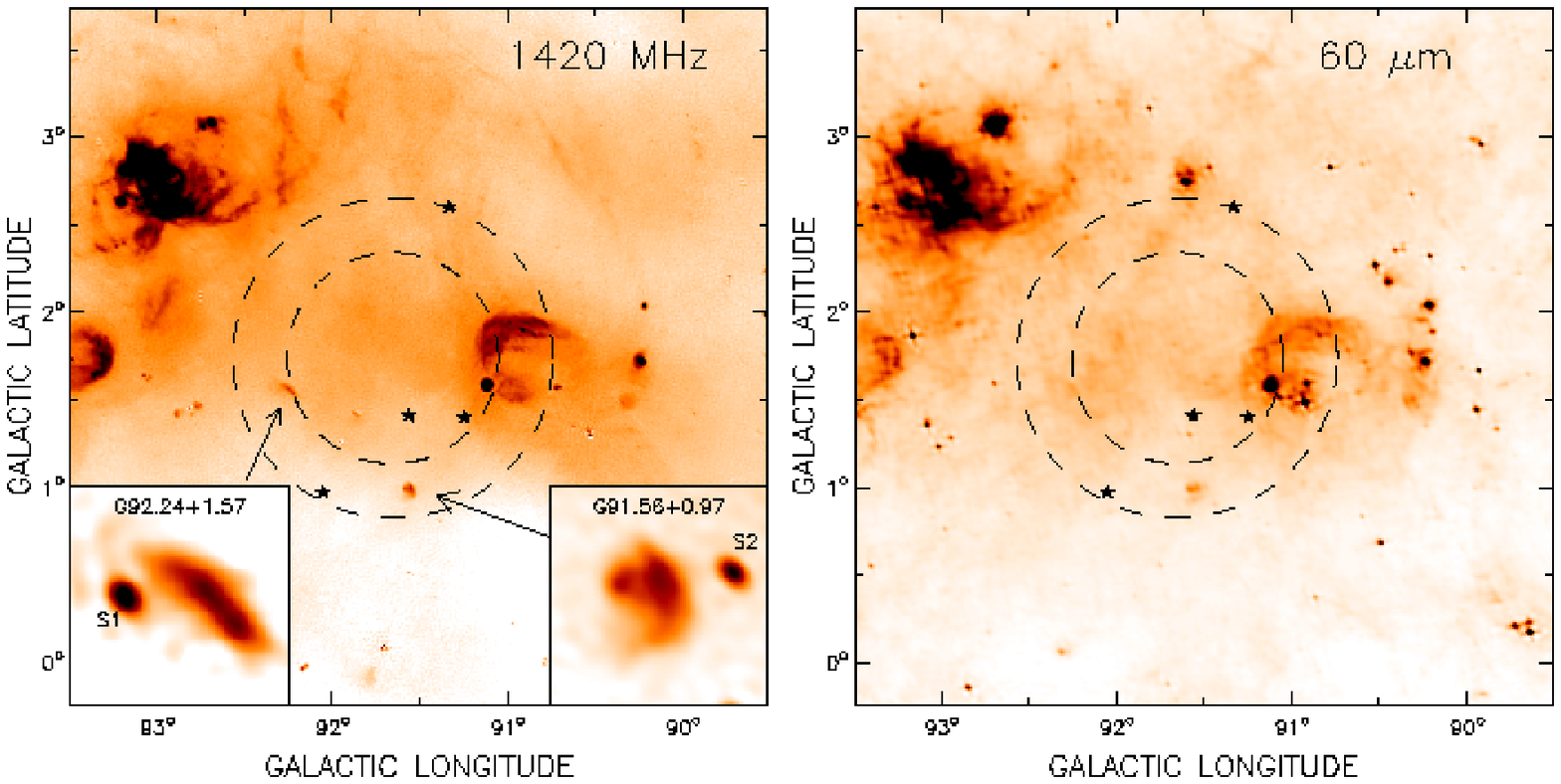}
\caption{1420 MHz radio continuum and 60 $\mu$m images from the CGPS
survey. Resolution for both is about 1\am. Point-like sources have
been removed from the radio continuum image.  Circles indicate
the approximate inner and outer boundary of the \hi\ shell seen in Figure \ref{average}. 
The two insets at the bottom of the radio image show a close-up view of G91.56+0.97
and G92.24+1.57 (point-like sources included, see text).}
\label{1420cont}
\end{figure*}

\citet{pin02} had already noted that no continuum (radio or IR)
counterpart seemed to be present.
Figure \ref{1420cont} shows the CGPS 1420 MHz radio continuum and 60
$\mu$m HIRES images of the same field of view as the preceding figures.
Discrete point sources have been removed from the radio continuum image
and the shadings chosen so as to highlight
the low-level emission centered on \G.
The relatively bright and large incomplete shell-like structure 
to the west of \G, near 
$(l, b) = (90\fdg 9, 1\fdg 6)$ is the \hii\ complex BG 2107+49 discussed
by \citet{van90} and located at a kinematical distance
of 10 kpc.  The circles delineate the approximate
inner and outer boundaries of the \hi\ shell,
revealing that there is no obvious radio continuum or
infrared emission related to \G.

The faint radio continuum emission seems to consist of filaments more or less 
aligned perpendicular to the Galactic plane, especially toward the north 
of the image, above
the opening in the \hi\ shell. The diffuse emission is more or less 
distributed homogeneously over the northern part of the image, although
there appears to be a slight excess of emission within the shell boundary,
in particular in its northernmost part.  In the south, the location where radio emission
becomes fainter corresponds roughly with the \hi\ shell boundary.

Two interesting and slightly extended ($6\mina - 8\mina$) structures 
(named G91.56+0.97 and G92.24+1.57) are seen near the inner
boundary of the \hi\ shell.  A close-up view of these structures is presented as
insets at the bottom of Fig.~\ref{1420cont}.  
These objects are discussed in more detail in Section \ref{trigger}.

We  also examined the CGPS polarization images (percentage polarization
and position angle) for this field and did not find any evidence for highly
structured emission associated with the shell or any of the slightly extended
structures discussed above.  This  agrees with the findings 
of \citet{uya03}, who concluded that all polarization structures observable in the
CGPS are generated and/or Faraday rotated closer than 2.5 kpc.

\section{Discussion}\label{discuss}

Some of the questions  we shall
try to answer are: can the observed \hi\ shell be a stellar wind shell
formed by the winds from one or more of the four candidate DWCL sources? If not,
what are the alternative shell-formation scenarios? Was the smaller 
\hi\ cavity (the one embedded in the wall of the larger \hi\ shell) 
caused by \IRdos\ and, if so, could \IRdos\ represent a
case of triggered star formation?
In this section we shall explore different possible formation scenarios based on the determined size, mass, age, and energy involved.
From here on we assume a distance of $15 \pm 3$ kpc for \G, as suggested 
initially by \citet{pin02}. 

 We note however that this distance is likely an upper limit.
Indeed, using a new method \citep{fos06} based on \hi\ column densities for the
determination of distances within the disk of the Galaxy, \citet{arv09} obtained
a new distance for the outer Galaxy \hii\ region CTB~102.  As
this object is at $l = 93\fdg115$, $b = 2\fdg835$, i.e.~very nearly along the
same line of sight as our shell, we can use \citet{arv09}'s
Fig.~5 to extrapolate their velocity-distance relation to obtain
a distance estimate of approximately 11.5 kpc for \G. This
slightly lower distance still corresponds to an object well in the
outer Galaxy at $R_g \approx 14.5 \rm\, kpc$  and $z \sim$ 400 pc.
Whenever possible, we shall show the distance dependence of simple parameters
by defining $d_{15} = d/15\,\rm kpc$.

\subsection{Parameters of the \hi\ shell}\label{param}

Error estimates for all the parameters are based on our ability to
estimate the spectral extent of the shell in velocity space, its 
spatial extent, and distance.

The \hi\ ring is clearly observed over some 14 \kms.  Following
the procedure described by \citet{pin98}, we derive a mass for the 
shell of $M_{\rm sh} = (2.3 \pm 1.3) \times 10^5 \, d_{15}^2 \, m_{\odot}$. 
Note that this value agrees with that obtained using 
\citet{hei84} equation for the mass swept up by a shell, namely 
$M_{\rm sh}( m_{\odot}) = 8.5 R_{\rm sh}^2 (\rm pc)$ which, taking
$R_{\rm sh} = (200 \pm 30)\, d_{15}$ pc, gives 
$M_{\rm sh} = 3.4 \times 10^5\, d_{15}^2 \,  m_{\odot}$.  Using our determined value for
the mass and considering a spherical cavity, we obtain for the initial (i.e. 
before the gas was swept up in the shell) neutral gas
particle density $n_o = (0.3 \pm 0.1) \, d_{15}^{-1}\,\rm cm^{-3}$.

Adopting an expansion velocity equal to half the velocity interval 
where the structure is observed, $v_{\rm exp}  = 7 \pm 2$ \kms,
we obtain for the kinetic energy stored in the expanding shell $E_{\rm kin} =
M_{\rm sh}\, v_{\rm exp}^2/2$  $= (1.1 \pm 0.6) \times
10^{50} \, d_{15}^2$ erg.  The kinematic age of the shell is given by $t ({\rm Myr}) =
\alpha\, R_{\rm sh}({\rm pc})/v_{\rm exp}$(\kms), where $\alpha =
0.25$ for a radiative SNR shell and $\alpha = 0.6$ for a supershell. 
We obtain $t = (7 \pm 2) \,d_{15}$ Myr ($\alpha = 0.25$) and $t = (17 \pm 5) \, d_{15}$ Myr
($\alpha = 0.6$).
With these determined observational parameters as a basis, we are  now
in a position to discuss the different formation scenarios for \G.

\subsection{Formation scenarios for the \hi\ shell}

In this section we assume for simplicity that the DWCL objects are located at the same
distance as \G, namely 15 kpc.  Of course we cannot rule out that there may be no physical 
association at all and that what we observe is simply a pure line-of-sight coincidence.

\subsubsection{The action of the DWCL source candidates}\label{dwcl}

The long kinematic timescale of 17 Myr, if \G\ has a SW origin,
suggests that more than one generation of massive stars
should be involved and/or that contributions from one or more SN explosion are to be expected.
Nevertheless, as a first step, it is of interest to estimate the contribution that the winds
of one or more of the DWCL candidates could make to the formation
of this large \hi\ shell.

We first calculated the total integrated IR luminosity of each candidate.
The derived  parameters are shown in Table \ref{colores}.
If, following \citet{coh95}, we assume the objects to be
WC8-9 stars, then their total luminosity is about \citep{cro07}
$L / L_{\odot} = 1.1 \times 10^5$. The IR luminosities thus represent
only about from 3 to 10 \% of the total stellar luminosity.  

Ignoring any intrinsic absorption by dust surrounding DWCL stars,
we may ask whether a WC8-9 star would be easily identified at a distance of 15 kpc
along the line of sight to \G.  
From the foreground \hi\ column density and
distance we can calculate typical values for the extinction and reddening of stars.
The measured foreground \hi\ column density is about
$1.1 \times 10^{22}\, \rm cm^{-2}$ , which corresponds to a reddening
of $E_{B -V} = 2.3$ mag ($E_{B -V} = N_{\rm H\,I}/ 4.8 \times
10^{21}\, \rm cm^{-2}$; \citet{boh78}). This gives a
visual absorption $A_v$ of 7.3 mag. Taking
for the absolute magnitude of WC8-9 stars \citep{cro07} $M_v = -4.5$, we
obtain $m_v = 18.7$.  A star of this magnitude would clearly
not stand out among field stars, although it would be
detectable. In their study of GSH\,90+03-99,
\citet{uya02} had estimated a
visual extinction of 6.4 mag toward {\it l} $\sim 90^{\circ}$ and
concluded that deep measurements were needed  to
detect early-type stars at a distance, for their object, of 13 kpc.

\begin{figure}
\resizebox{\hsize}{!}{\includegraphics[width=8.5cm,angle=-90]{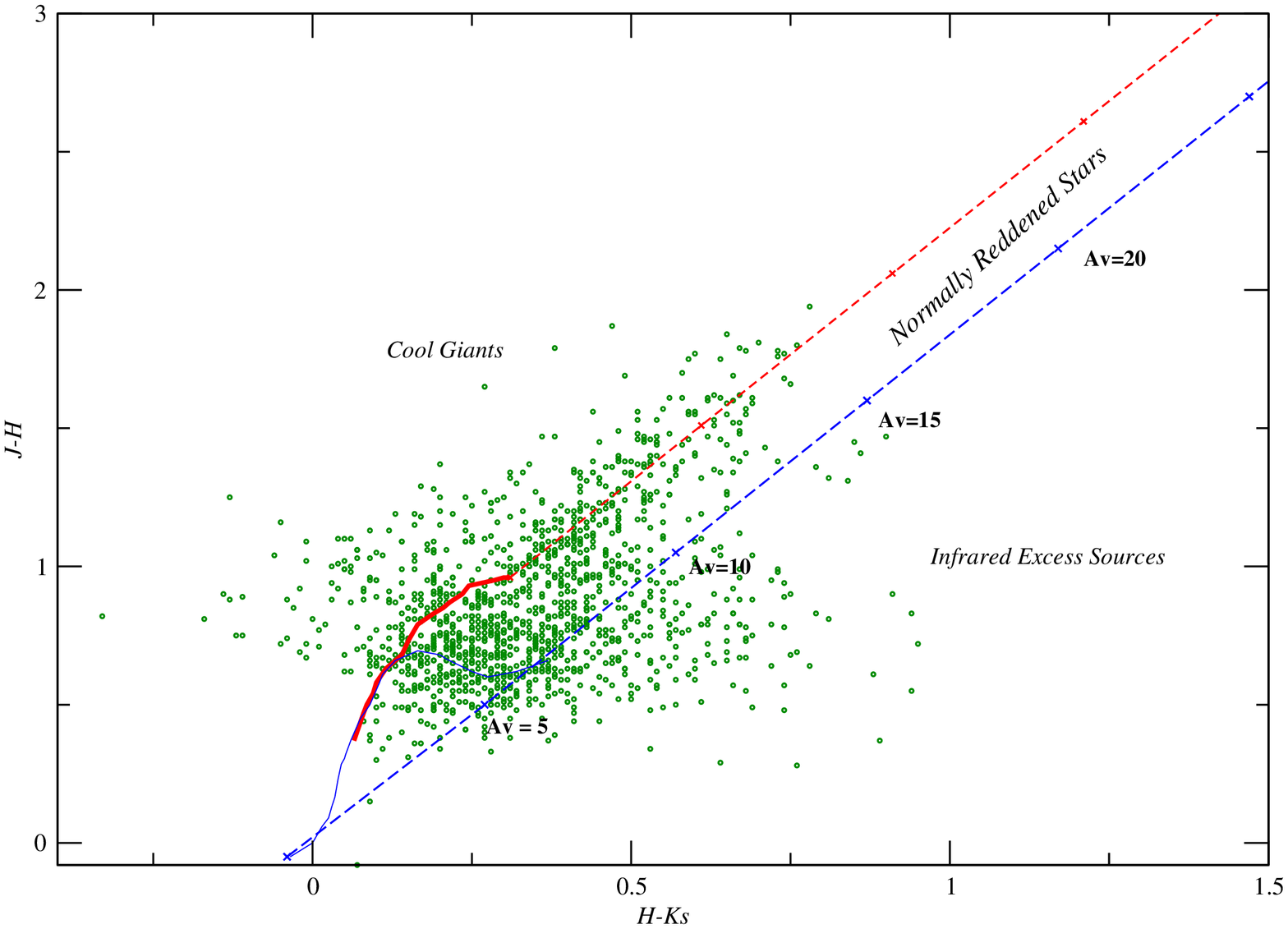}}
\caption{
$(H-K, J-H)$ color-color diagram of the 2MASS sources found in a 400\as circular 
area centered at ($l, b$) = (91\fdg7, 1\fdg83).  The two solid curves 
represent the location of the main sequence (blue line) and the giant stars (red line) 
derived from \citet{bes88}. The parallel dashed lines are reddening vectors 
with the crosses placed at intervals corresponding to five magnitudes of visual 
extinction. We assumed the interstellar reddening law of \citet{rie85}, 
$A_J/A_V = 0.282$ and $ A_H/A_V = 0.175$.
}
\label{2mass}
\end{figure}

There are reasons however to believe that the value of absorption derived
from the \hi\ column density is a significant underestimate of the true
value.  As an alternative, we used
the 2MASS Point Source Catalogue \citep{skr06} to produce a $H-K_s$ vs $J - H$ diagram
(Figure \ref{2mass}).  There are 1573 2MASS sources in a
circular area centered at ($l, b$) = ( 91\fdg7, 1\fdg83) within a radius
of $400 \seca$. These sources are indicated by green dots in Fig.~\ref{2mass}.
The positions of the dereddened early-type main sequence and giant
stars are indicated with blue and red solid lines, respectively. The
blue and red dashed lines show the reddening curve for  O9 V and M0 III
stars, respectively.  Thus, normally reddened main sequence stars lie
between the two dashed lines. Because absorption is proportional
to distance, we can infer that statistically the most distant sources have a visual absorption of about 13
mag along this direction in the Galaxy. Hence, as we are dealing with a structure located in the outer part
of the Galaxy at about 15 kpc, it is likely that the visual absorption is
significantly higher than 7.3 mag. If absorption is as high as 13 mag, the expected apparent magnitude of WC8-9 stars could be as faint
as $m_v = 24.4$.  The higher absorption derived
from the color-color diagram suggests the presence of molecular gas 
along the line of sight in this direction. 
It also makes the optical identification of any massive star at that distance 
even more problematic.

 Note that lowering the distance to 11.5 kpc only makes the above
magnitudes brighter by 0.6 mag.

Next we estimate the energy injected by the stellar winds, 
$E_w = \dot{M} v_w^2\, t /2 $.  Adopting \citep{cro07} for a single WC8-9  a mass loss rate of
$\dot{M} = 10^{-5}\, m_{\odot}\, \rm yr^{-1}$,
appropriate for a solar metallicity $Z_{\odot}$, 
and a wind velocity of $v_w = 1500$ \kms, we obtain 
$E_w = 2.25 \times 10^{44}\, t$(yr) erg. If the WC
phase lasts $\sim 10^5$ yr,  each star would impart $E_w =
2.25 \times 10^{49}$ erg to its local ISM during this phase. But only a fraction $\epsilon$
of this energy gets transferred to the gas.  According to
evolutionary models of interstellar bubbles, the expected energy
conversion efficiency $\epsilon = E_{\rm kin}/E_w$
is on the order of 0.2 or less \citep{mcc83}. Observationally, the values of $\epsilon$ derived from optical
and radio observations can be as low as 0.02 \citep{cap03}, indicating
that there are cases where severe energetic losses may occur.

Furthermore, given the location of \G\ in the outer Galaxy, 
the effect of metallicity should be taken into account. 
\citet{nug00} suggested that the mass-loss rate depends on 
metallicity $Z$ as  $\dot{M} \propto Z^m$, with  $m = 0.5$.
The velocity of the wind is also lower at lower metallicity \citep{nug00}.
\citet{sti01} concluded that the wind energy $E_w$ decreases by a 
factor of about 3 when $Z$ decreases by a factor of 10 for stars 
with the same luminosity.
Thus, for $Z = 0.1\,Z_{\odot}$, each WC8-9 star would simply inject 
$E_w \sim 7.5 \times 10^{48}$ erg in the ISM during the WC phase. This
implies that even with a relatively high efficiency $\epsilon = 0.2$,
the injected kinetic energy is only $E_{\rm kin} = \epsilon\,E_w =
1.5 \times 10^{48} \rm\, erg$.
Summarizing, neither for $Z = Z_{\odot}$ nor 
for $Z = 0.1 \,Z_{\odot}$ could the \hi\ shell
have been created {\it only\/} by the stellar winds 
of the DWCL sources in their WC phase. Even by considering a comparable
contribution from the previous evolutionary phase of each star, the available kinetic
energy falls considerably short of the observed shell kinetic energy $E_{\rm kin} = 
(1.1 \pm 0.6) \, \times 10^{50} \,d_{15}^2\, \rm erg$.

\subsubsection{The action of unseen massive stars}\label{massive}

We have failed in finding any massive star in the area.  However we have
seen in Section \ref{dwcl} that distance and absorption conspire to
make even bright early-type stars inconspicuous.  At a distance of 15 kpc
and with an absorption of 7.3 mag, based on the \hi\ column density, 
an O3V star ($M_v = -6.0$) would have
$m_v = 17.2$, whereas a B1V star ($M_v = -3.2$) would have $m_v = 20.0$.
For an absorption of 13 mag, derived from the 2MASS color-color diagram, the
corresponding magnitudes would be 22.9 and 25.7, for an O3V and a B1V
star respectively.   As mentioned above, a shorter distance of 11.5 kpc
leads to estimated magnitudes brighter by only 0.6 mag.
Although one or more early-type star cannot be easily detected
at the distance of \G, we cannot rule out the presence of these stars
and their possible role in forming the \hi\ shell.  Furthermore, given that
massive stars evolve in a short time, the originating stars might have
vanished a long time ago and be not observable any more.

If the origin of \G\, is indeed the action of many massive stars, 
from the estimated value of the kinetic energy of the shell ($E_{\rm kin} = (1.1 \pm 0.6) \times
10^{50}$ erg, see Section \ref{param}) we can estimate how many massive stars would have been 
needed to create it. Adopting mean stellar wind parameters for O-type stars 
(i.e. $\dot{M} = 2 \times 10^{-6}$ m$_\odot$\,yr$^{-1}$ and  $v_w = 2000$ \kms, \citet{mok07}) 
and considering that the main sequence phase lasts for at least $3 \times 10^6$ yr, 
the injected wind stellar energy is about $1.4 \times 10^{50}$ erg.
Bearing in mind that the conversion efficiency is about or less than 0.2 and considering the 
possible effects of a lower metallicity environment, the energy released by the stellar winds 
should be greater that $5.5 \times 10^{50}$ erg. Thus, at least four O-type stars were
necessary to create \G.
 In the case of a lower distance of 11.5 kpc, the energy released by the stellar winds should be greater than $3.2 \times 10^{50}$ erg, implying that at least three O-type stars were required.

\subsubsection{A supernova explosion}\label{snexplo}

\citet{uya02} proposed and interesting procedure for setting
various constraints on an SNR origin, which they applied to their \hi\
shell and another one previously discovered by \citet{sti01}.
Their procedure essentially rests on the
expression giving the maximum observable radius of an SNR before it merges
with the ISM \citep{cio88}:
$$R_{\rm merge} = 51.3 \, E_0^{31/98} \, n_0^{-18/49}\, Z^{-5/98}\rm\, pc,$$
\noindent where $E_0$ is the explosion energy in units of $10^{51}$ erg,
$n_0$  the mean ambient particle density in $\rm cm^{-3}$, and
$Z$  the metallicity normalized to the solar value. 
The input parameters to the above equation are the observed radius and particle density,
$R_{\rm sh} = 200 \,d_{15}\,\rm pc$ and $n_0 = 0.3 \, d_{15}^{-1} \,\rm cm^{-3}$.
The results for \G\ are shown as the last entry
in Table \ref{snrparameters},  where the numerical values are for $d = 15 \,\rm kpc$.
The first two entries are identical to Table 2 of \citet{uya02}
(with the addition of two parameters and a slight change in notation for added clarity).

\begin{table*}
\caption{Constraints on a single SNR origin}
\label{snrparameters}
\centering
\begin{tabular}{lccccccl}
\hline\hline
Object  & $n_0$ & $R_{\rm sh}$ & $Z$ & $R^*_{\rm merge}$ & $E^*_0$ &   $d_{\rm max}$ & Reference\\
 & $\rm (cm^{-3})$ & (pc)  & $(Z_{\odot})$ & (pc) & ($10^{51}$ erg) & (kpc)& \\
\hline
GSH\,138--01--94  & 0.27 & 180 & 0.1 & 93.3 & 8.0 & 5.9 &  \citet{sti01}\\
             & & & 1.0 & 83.0 & 11.6 & 4.9  & \\
GSH\,90+03--99 & 2 & $55\times 110$ & 0.1 & 44.7 & 6.3 & 5.2 & \citet{uya02}\\
  & & & 1.0 & 39.8 & 9.1 & 4.3  & \\
\G & 0.3 & 200 & 0.1 & 90 & 12.6 & 6.9 & This paper\\
  & & & 1.0 & 80 & 18.2 & 6.1 & \\
\hline
\end{tabular}
\begin{list}{}{}
\item $R^*_{\rm merge}$ (column 5): merging radius for $E_0 =1$.
\item $E^*_0$ (column 6): value of $E_o$ required for $R_{\rm merge}$ to equal the
observed radius.
\item $d_{\rm max}$ (last column): 
maximum distance for a single SN explosion with $E_0 =1$.  

\end{list}
\end{table*}

The column labelled $R^*_{\rm merge}$ (column 5) gives the merging
radius assuming a canonical value of $E^*_0 =1$ for the explosion energy.
As with the other two shells, the observed radius  $R_{\rm sh} $
is about twice as large as the predicted merging radius $R^*_{\rm
merge}$, which means that the shell should have vanished well
before reaching the observed radius. This implies that either $E_0$ is
larger and/or the object is closer. 
The column labelled $E^*_0$ (column 6) gives the energy required for $R_{\rm merge}$ 
to be equal to the observed radius of the shell.
The morphology of \G\
is clearly too regular and well defined to represent a merging shell
so that the value of $E^*_0$ is a lower limit.  
The last column is the
maximum distance \G\ would have to be in order to be the result of a single SN 
explosion with $E_0 =1$.   This distance corresponds
to a systemic velocity of about $-52$ \kms\ which, as 
is the case for the other two
shells, is totally unrealistic, given the observed systemic velocity
of $-114$ \kms.

If we compare the three shells, we see that the first and last one are
very similar in radius and mass, because they are virtually at the same distance
and evolving in an ISM of comparable particle density.  The second one,
GSH90+03-99, is somewhat closer and appears to be located in a significantly
denser medium.  Nevertheless, as discussed by \citet{uya02},
neither of the three shells could have been caused by a single SN explosion.
Furthermore, as emphasized by \citet{uya02},
this argument applies as well to a single O- or B-type stellar wind bubble.
 Using 11.5 kpc as the distance would only change $R^*_{\rm merge}$ and
$d_{\rm max}$ by about 10\% and $E_o^*$ by about 40\%, thus not altering the
general conclusion.

\subsection{Additional constraints from radio continuum emission}\label{adcons}

From Figure \ref{1420cont} it is clear that there is no evidence of enhanced
radio continuum emission associated with \G.  Although this absence is
somewhat puzzling for \hi\ shells associated with known optical
objects such as WR or Of stars, it is less so here. The kinematic age of
the shell could be as high as 17 Myr and massive O stars and early B stars
could well have ended their lives.  However, because of the large distance
and large extinction involved, we cannot completely rule out the presence of
one or more of these massive stars.  This statement can however be made
more precise.

Bellow we shall use the following parameters for the \hi\ shell
(note that $1\mina$ is equivalent to 4.5 pc at a distance of 15 kpc):
inner and outer radius, $R_i = 160 \rm\, pc$, $R_o = 200 \rm\, pc$, shell
thickness $\Delta R = 40 \rm\, pc$ and initial particle density 
$n_o = 0.3 \rm\, cm^{-3}$.  These are average quantities because obviously 
the shell radii and thickness vary significantly with azimuthal angle.
With these parameters, we deduce that the particle density within the shell
is now $n_s = 2.6 \,\times\, n_o = 0.8 \rm\, cm^{-3}$. For simplicity, we
shall neglect the presence of helium in the following simple estimates.

\subsubsection{The simplest case: a homogeneous shell}

We assume that an ionizing star is located at the center of the \hi\ shell and
ask to what extent the inner side of the \hi\ shell should be ionized, assuming that
negligible material is present inside the shell.  This
is the classical Str\"omgren sphere problem where, instead of integrating from
0 to some maximum radius $R_m$, we integrate from $R_i$ to $R_m$:
$\int_{N_u}^0 {\rm d}S_u (r) = -\int_{R_i}^{R_m} 4\,\pi\,r^2\,n_e^2\,\alpha^{(2)}\, 
   {\rm d}r$,
where $S_u (r)$ is the rate of ionizing photons at distance $r$, $N_u$ is
the total number of ionizing photons emitted by the star,
$n_e$ the electron number density (assumed equal to $n_s$) and $\alpha^{(2)}= 
3.09 \times 10^{-13} \rm cm^{3}\,s^{-1}$ is the recombination coefficient excluding captures to the
ground level.
Solving for $R_m$, we obtain

\begin{equation}
R_m = R_i\,(1 + {\cal R}/R_i)^{1/3}, \quad {\cal R} = {{3\,N_u}\over
 {4\,\pi\,n_e^2\,\alpha^{(2)}\,R_i^2}}.
 \label{rmaxeq}
\end{equation}

If $R_m > R_o$, then the \hi\ shell is fully ionized (and should not exist!)
and some photons escape freely.  For our parameters, 
${\cal R} = 0.19\,N_{u,47}\,\rm pc$, where $N_{u,47}$ is the ionizing
photon rate in units of $10^{47}\,\rm s^{-1}$.
The total number $N_u$ of UV ionizing photons also fixes the total
observed flux density $S_{\nu}$ at a given frequency
and is given by \citep[e.g.,][]{rub68,cha76}
\begin{equation}
   N_u  = 0.76 \times 10^{47}\,\,T_4^{-0.45}\,\nu_{\rm GHz}^{0.1}
             \,S_{\nu}\,\,d_{\rm kpc}^{2},
\end{equation}
where $T_4$ is the electron temperature
in units of $10^4$ K, $\nu_{\rm GHz} = 1.42$ the frequency in GHz
and $S_{\nu}$ is in Jy.
Using $T_4 = 0.7$ and $d_{\rm kpc} = 15$, we obtain
$S_{1.42} = 4.8 \, N_{u,47}\rm\, mJy$. This
estimate for $S_{\nu}$ is an upper limit, because photons can escape through
inhomogeneities of the shell (on scales smaller than the beam) or
through the opening in the north.

\begin{table}
\caption{Constraints from radio continuum$+$}
\label{radcont}
\centering
\begin{tabular}{lrrrrr}
\hline\hline
Stellar type  & $N_{\rm u,47}$  & $S_{1420}$  & ${\cal R}$ &  $R_{\rm m}$ \\
 & ($\rm s^{-1}$) & (mJy) & (pc) & (pc)\\
\hline
O3V & 741 & 3580 & 141 & 198\\
O6V & 219 & 1060 & 42 & 173\\
B0V & 14.5 & 70 & 3 & 161\\
\hline
\end{tabular}
\begin{list}{}{}
\item{$^+$} For a distance of 15 kpc.
\end{list}
\end{table}

Table \ref{radcont} shows the results of these simple calculations for three
different stellar types, chosen to illustrate different possibilities. Evidently the observations are totally incompatible with the presence of an O3V
star (or any star with the same value of $N_u$) as the \hi\ shell would be
essentially totally ionized.  In order for the neutral shell to be actually present,
there would then have to be a substantial amount of ionized gas {\em within\/} $R_i$
to absorb the ionizing photons.  This
goes against the canonical view of a shell or supershell as surrounding an
essentially empty cavity. Figure \ref{1420cont} also fails 
to show any significant amount
of continuum emission in excess of the diffuse emission seen in this general
direction which, given the large distance of the shell, is likely to be
foreground emission.

At the other extreme, a B0V star (or any star with  
$N_u \le 1.5\,\times\,10^{48}\,\rm s^{-1}$ would ionize a negligible part of
the \hi\ shell, here hardly 1 pc.  
We can use eq.~(\ref{rmaxeq}) to estimate the ionizing photon
rate, which would be enough to ionize a thin layer 1 beam size 
in thickness ($1\mina$ or about 5 pc). This gives $\log N_u = 48.90$, corresponding
to an O8.5V star.  Note that WR stars have $\log\, N_u$ in
the range 48.6 to 49.4 (Crowther 2007).

As for the intermediate case of an O6V star, Table \ref{radcont} shows that it
would ionize about 13 pc of the inner \hi\ shell.  This ionized gas would have
an angular thickness of about $3 \mina$ and a flux density of about 1 Jy.
An {\em upper\/} limit to the brightness temperature $T_B$ can be obtained by
assuming this flux to originate from an annulus $35 \mina$ in radius and
$3 \mina$ in thickness (giving an area of about 660 arcmin$^2$ or 390 CGPS
beam areas at 1420 MHz).  We obtain $T_B < 2.3 \rm\, mJy/beam \approx 0.4\rm\,K$
or, given the measured noise of 0.082 K, about $5\,\sigma$.  An alternative means
of estimating $T_B$ is to use an approximate emission measure given by $EM = \int
n_e^2\,{\rm d}l \sim n_e^2\,R_o \sim 100 {\rm\,pc\,cm^{-6}}$, from which
the brightness temperature is $T_B = (EM/566{\rm\,pc\,cm^{-6}}) 
\rm \, K = 0.17\rm\, K$.  Given
the spatially varying continuum surface brightness within the inner shell boundary,
such a faint ionized layer would not be easily detected by our observations.

Summing up, if the \hi\ shell is relatively homogeneous, which is
far from certain, we can rule out
the presence of any star more luminous than O6V or with more than about
$2\times 10^{49}\rm\  ionizing\ photons\ s^{-1}$.  Placing the shell at the
slightly closer distance of 11.5 kpc would have the effect of making it
slightly thinner and closer to any central star with the result that for
any given spectral type, more of the shell would be ionized.

\subsubsection{A clumpy shell}

In addition to an obvious opening in the north, the \hi\ shell does appear
to be inhomogeneous on scales comparable to the beam size (about $1 \mina$ or
4.5 pc at a distance of 15 kpc).  If it were inhomogeneous on scales smaller
than the beam (thus undetectable in the CGPS image), the \hi\ shell could
let a significant number of photons escape, thus leading to a more diluted and
fainter ionized layer.

The existence of inhomogeneities, whether they are called clumps, filaments, clouds,
or cloudlets, has been invoked in a number of contexts.
As a possibly extreme case (with
regards to size), \citet{eve10} recently 
postulated the existence, within the 
stellar wind cavity of the $2\farcm2$ nebula N49, of a few hundred cloudlets
of internal density $10^5 \rm\,cm^{-3}$ and of a radius as small as 0.05 pc.
These would provide the dust needed to explain
the 24 $\mu$m emission observed within the N49 cavity through erosion and evaporation.

We have no way of knowing the degree or type of inhomogeneities which could
be present, however, for illustrative purposes, we consider inhomogeneities in
the form of clumps with a radius $r = 1\rm\,pc$ and internal density 
$n^{\prime} = 10^2 \rm\,cm^{-3}$, thus each clump would have a mass $m = 10.4\,m_{\odot}$.
With a total \hi\ shell mass of $2.3 \times 10^5 \,m_{\odot}$, this implies
that there are $N=2.2 \times 10^4$ such clumps.  The volume filling factor
is $f \approx (4 \pi r^3N/3)/(4\pi R_i^2\Delta R) = r^3N/(3 R_i^2\Delta R)
\approx 7.6\times 10^{-3}$.  To estimate the transparency of the shell, we note
that as viewed from the center of the shell, the clumps cover a fraction $\epsilon$
of the shell area where $\epsilon \approx N \pi r^2/(4 \pi R_i^2) =
3\, f \Delta R/(4r) \approx 0.23$.  

In other words, about 77\% of the ionizing photons would escape through the
shell or, taking the case of an O3V star, only $168 \times 10^{47} \rm\ 
ionising\ photons\ s^{-1}$ could contribute to partially ionize the clumps 
from $R_i$ to $R_o$, resulting in an undetectable radio continuum surface
brightness.

In summary, if the \hi\ shell is basically homogeneous, there can be
no star producing more than about $2\times 10^{49}$ ionizing
photons per second.  Depending on the degree of clumpiness of the shell however,
such powerful stars could still be present.  Nevertheless, given the possibly
large kinematical age of the shell (up to 17 Myr), it is most likely that there
are no O stars left inside the shell and that any B star, if present, contributes
negligibly to ionization.

\subsection{Triggered star-formation?\label{trigger}}

Shocks in expanding supershells are widely believed to be the primary mechanism for
triggering star formation \citep{elm98}.  Shells behind shock fronts experience  
gravitational instabilities that may lead to the formation of large condensations
inside the swept-up material, and some of these may produce new stars \citep{elm98}.
An increasing body of observational evidence confirms the importance of this mechanism
\citep[e.g.][]{pat98, oey05,arn07, cic09}.
Given the size of \G, it seems reasonable to expect some signs of recent star-formation
activity in or near the shell border.

\subsubsection{Radio continuum sources}

The 1420 MHz radio continuum image
(Fig.~\ref{1420cont}) shows that a few extended, yet relatively
compact objects are found near the inner periphery of the \hi\ shell.
Apart from G91.11+1.57, the so-called
``head'' of the \hii\ complex BG 2107+49 discussed by \citet{van90},
there are only two such objects within the boundary of the \hi\ shell.
Could these be \hii\ regions whose formation was triggered by the
expanding \hi\ shell? An unambiguous answer to this question requires a
determination of both the distance and the radio spectral index.

We can 
obtain an estimate of the spectral index 
$\alpha$ ($S_{\nu} \propto \nu^{-\alpha}$)
by measuring the flux densities at 1420 and 408 MHz. 
Yet for both G91.56+0.97 and G92.24+1.57, a substantially
bright compact radio source  
is present in their immediate vicinity (named  $S_1$ and $S_2$, see insets
of Fig.~\ref{1420cont}).  Whereas this poses no problem at 1420 MHz, the larger
beam at 408 MHz results in a partial blend of these point sources with the
nearby extended structure, particularly severe for G92.24+1.57.

To obtain the spectral indices, we first subtracted all point-like sources from
the images at both frequencies with a two-dimensional
Gaussian-fitting routine (Fig.~\ref{1420cont} 
is the result at 1420 MHz).  Because both   $S_1$ and $S_2$ were successfully subtracted at
this frequency, we then used the {\sc imview} program\footnote{This program is part of
the DRAO export software package} to measure the flux density of the remaining
extended source.  The error was estimated by using slightly different background levels.

On the 408 MHz image however, the larger beam size ($\approx 4^{\prime} \times 3^{\prime}$)
results in the Gaussian-fitting routine which finds
both  $S_1$ and $S_2$ to be extended, in contradiction with the fact that both are
unresolved even at 1420 MHz.  Inspection of the 408 MHz image showed however that the
program did successfully subtract the {\sl combined} flux density of the extended structure
{\sl plus\/} the nearby compact source, for  both G91.56+0.97 and G92.24+1.57.  
In order to obtain
the separate flux densities, we proceeded in two stages.  Firstly, from the point source
flux density at 1420 MHz, we estimated the 408 MHz flux density, starting with a trial
spectral index $\alpha = 0.75$, which is representative of the spectral index of extragalactic
radio sources between 178 and 1400 MHz \citep{pac77}.  We then created a Gaussian source
with this flux density with the 408 MHz beam size that we removed from the original
image.  We varied $\alpha$ until
the best artefact-free subtraction was found.  The flux density of the extended
structure was then estimated as the difference between the combined flux density and that
of the nearby point source.   The 408 MHz flux densities of the two point sources, $S_1$
and $S_2$, were cross-checked by comparing our 408 MHz values with the 327 MHz values from the
Westerbork survey \citep{wes327}.  The values for $S_1$ and $S_2$ at 327 MHz are 57 mJy
and 90 mJy, respectively, in satisfactory agreement with our determinations.

Table \ref{fluxalpha} summarizes the obtained flux densities and spectral indices,
together with some independent measurements.  Both sources  $S_1$ and $S_2$ have a non-thermal
spectrum consistent with their origin as extragalactic radio sources. 
 As for G91.56+0.97 and G92.24+1.57,
despite a few discrepant measurements, probably arising from a difference in
background removal and/or inclusion of nearby point sources,
they show a spectral index consistent with a thermal nature.

Without a distance estimate it is impossible to ascertain 
whether they are associated or not with the \hi\ shell, however 
given their position on or near the \hi\ shell, the possibility that
they might be \hii\ regions
whose formation was triggered by the expanding \hi\
shell is worth pursuing.

\begin{figure}
\resizebox{\hsize}{!}{\includegraphics[width=8.5cm]{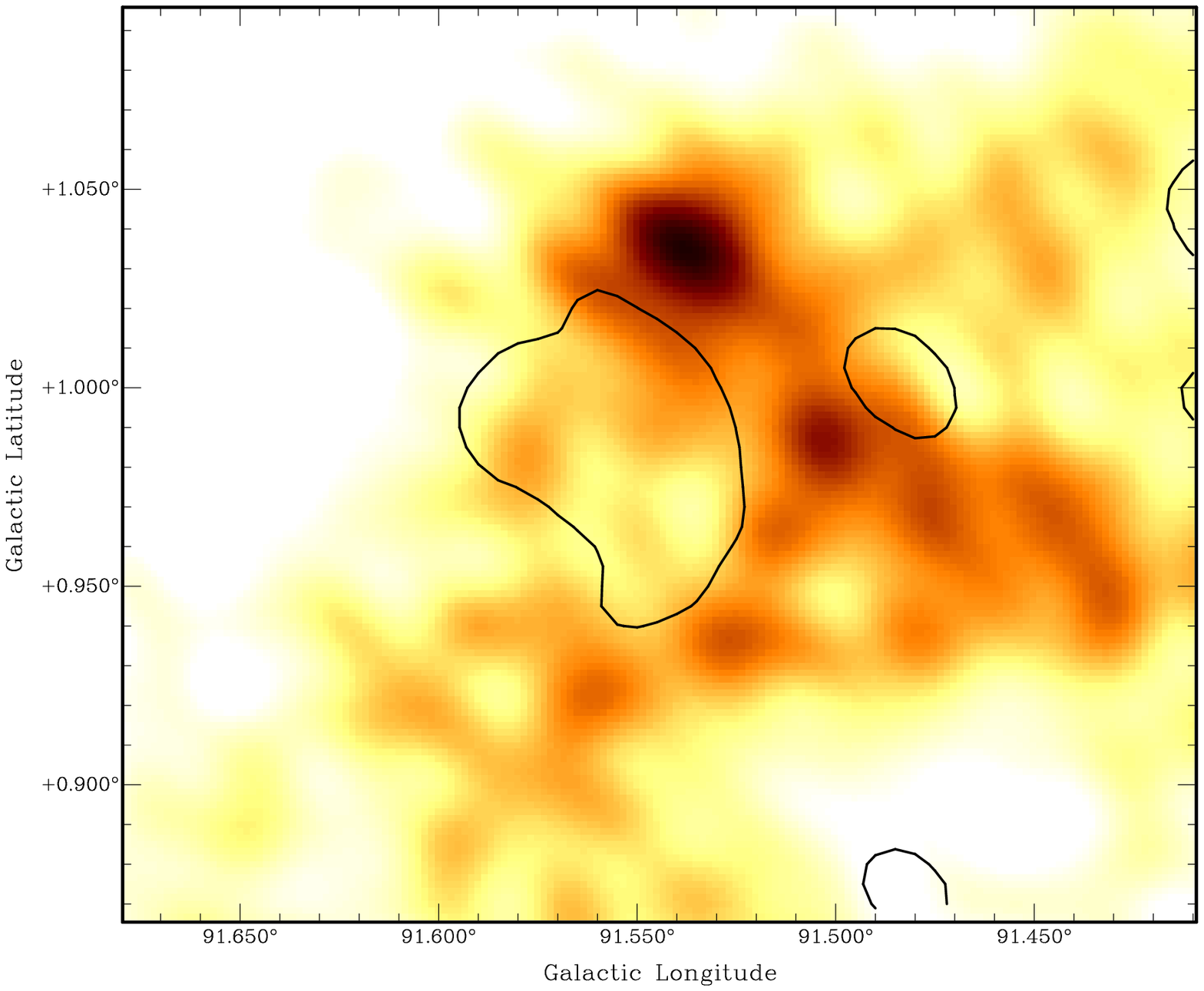}}
\caption{\hi\ emission distribution averaged between  --46.0 and --49.3 \kms.The black contour corresponds to the 8.5 K level of the 1420 MHz radio continuum emission.
}
\label{fuente1-hi}
\end{figure}

Both sources are unfortunately 
too faint for a significant
absorption spectrum to be obtained, which would enable us to set a
limit on the distance.
Another way to try to estimate their distances is to look for signatures in the \hi\ emission distribution around these sources (since unfortunately no molecular data is available for this region).
An inspection of the entire \hi\ data cube shows a minimum in the velocity range from about --46.0 to --50.0 \kms\, in the area of G91.56+0.97. A well defined arc-shape structure of enhanced emissivity surrounds  the minimum toward lower galactic longitudes. 
In Fig. \ref{fuente1-hi} the contour delineating the \hii\ region \mm\ is shown superimposed on the \hi\ emission distribution averaged in the velocity interval mentioned above.
The excellent morphological correlation observed between both structures would put \mm\ at a distance of about 6.5 kpc, implying that it is not related to \G.
As for \nn\, we did not find any clear \hi\ structure that could be associated with this source.

As mentioned previously, the source G91.11+1.57, associated with
the \hii\ complex BG 2107+49 (van der Werf \& Higgs 1990), also lies near the periphery
of the large \hi\ shell.  Could the two be associated?
\citet{hig87} obtained an H103$\alpha$ recombination line spectrum,
which showed this source at a velocity of $-79.5 \pm 1.1$ \kms. The resolution of
their NRAO 43-m observations was 5\am.  
Van der Werf \& Higgs (1990) obtained a similar value of$-80 \pm 1$ \kms, 
using DRAO \hi\ absorption spectra and an H112$\alpha$ recombination line 
spectrum obtained with the  100-m Effelsberg telescope at a resolution of
$2\farcm 6$. These observations would seem
to place G91.11+1.57 and our \hi\ shell at significantly different distances.
However, the VLA 5\as\ radio continuum image of Higgs et al.~(1987) 
shows the head of BG 2107+49 to consist of a set of discrete knots of
emission about 10\as\ in size and a more extended and diffuse region about 3\am\ in diameter.
Although we cannot completely rule out the possibility that some of the knots
could be at the larger distance of our \hi\ shell, the fact that
the H103$\alpha$ recombination line spectrum of Higgs et al.~(1987)
is extremely well fitted by a single line at $-79.5$ \kms\ with essentially no
residuals near $-114$ \kms\ would seem to rule this out.

\subsubsection{Infrared sources}

We feel that the case for triggered formation is stronger
for \IRdos.   This IRAS point source is seen projected inside the
shell wall and appears itself surrounded by a smaller \hi\ bubble
(Fig.~\ref{average}).
This smaller \hi\ cavity resembles the ones found by \citet{dub90}
and \citet{arn07}.  In both cases the authors found a small cavity
immersed inside a larger HI shell.  However, their interpretation about the
origin of the  structures differed.
\citet{dub90} concluded that both the large HI shell and the smaller bubble
had been created by the same star, HD\,197406, during different stages of its evolution.
On the other hand, \citet{arn07} found an HI cavity  around the OB association
Bochum 7 and concluded that this association might have been born as a consequence of
the evolution of the large shell GS263-02+45.

From Fig.~1 we can infer that the size of the smaller cavity is about 0\fdg 3.
At the distance of \G, this implies a linear size of about 80 pc.
Given that this cavity can be observed over at least 10 \kms\, (see Fig.~2),
a  lower limit for the expansion velocity  can be assumed as 5 \kms. 
Under these conditions we derived an  upper limit of about 5 Myr for the 
dynamical age  if a stellar wind origin is considered.
This age is significantly smaller than the one obtained for the large shell (7 - 17 Myr),
but larger than the duration of the WC phase ($10^5$ yr), suggesting that the O-phase
of the current WC star would also have contributed to its formation. 
Taking into account that progenitors with a mass of some 40 to 50 $M_{\odot}$ 
are suggested for late WC stars \citep{cro07}, 
the dynamical age estimated for the small cavity is consistent  
with the time that these stars stay on the main sequence, between 3.7 and  
4.9 Myr \citep{sch92}.
Based on the age estimate difference,  the most probable scenario is one where the
large shell, \G, was created by the joint action of several massive stars and where
at least one of them has already exploded as a supernova.  
Furthermore, as the shell evolved,
new stars may have been triggered in its dense border, 
IRAS\,21147+5016 being one of them.
It is of interest that this smaller bubble (Fig.~\ref{1420cont}) appears to
be opened on the exterior side of the larger \hi\ shell.   This would be
expected for a star whose formation would have taken place toward the
exterior of the expanding shell, leading it to first burst outwards.

\begin{table*}
\caption{Flux densities and spectral indices of compact sources$^+$
($S_{\nu} \propto \nu^{-\alpha}$)}
\label{fluxalpha}
\centering
\begin{tabular}{rlccccrrl}
\hline\hline
\# & Object & $F_{1420}$(mJy) & $\sigma_{1420}$(mJy) &  $F_{408}$(mJy) & $\sigma_{408}$(mJy)
         & $\alpha$&  $\Delta\, \alpha$  &  Notes$^a$\\
\hline
1 &$S_2$           & 37 & 2    & 76 & 9 & 0.58 & 0.10 & This paper, see text\\
2 &G91.56+0.97 & 191 & 19 & 159 & 27 & $-0.15$ & 0.16 & This paper, see text\\
3 &            & 261 & 11 & 346 & 28 & 0.09& 0.06 & vdWH, HvdW$^{\mathrm{b}}$ \\
4 &            & 238 & 9.5 & 79.2 & 1.1 & $-0.88$ & 0.03 & Kerton et al.~(2007)\\
\hline
5 & $S_1$           & 25 & 2 & 59  & 4 &  0.69  & 0.08 & This paper, see text\\
6 &G92.24+1.57  & 132 & 10 & 104 & 15 & $- 0.19$ & 0.13 & This paper, see text\\
7 &             & 99.4 & 6.9 & 168 & 31 & 0.42 & 0.16 & Kerton et al.~(2007)$^{\mathrm{c}}$\\
8 &             & 146 & 22 & -- & -- & -- & -- & vdWH\\
\hline
\end{tabular}
\begin{list}{}{}
\item[$^+$] Sources  $S_1$ and $S_2$ are the nearby point sources seen in the
insets of Fig.~3
\item[$^a$] HvdW = Higgs
   \& van der Werf (1991), vdWH = van der Werf \& Higgs (1990)
\item[$^{\mathrm{b}}$] Flux density at 408 MHz includes
 contribution from source $S_2$ (29P62 at 1420 MHz -- HvdW).
Note that the spectral index listed there comes from a linear fit
using measurements at five frequencies between 408 MHz and 4.85 GHz
\item[$^{\mathrm{c}}$] Likely includes source  $S_1$ at 408 MHz
\end{list}
\end{table*}

\section{Conclusions}\label{concl}

The measured kinetic energy of expansion of \G\ is far too large to have
been produced solely by the action of one or all of the four DWCL sources
possibly associated with the \hi\ shell.  That the shell is
very symmetrical whereas the DWCL sources are located well off-center
suggests that if they are indeed inside the shell, their shaping
influence is minimal.  We have found no evidence for
the presence of other massive stars, but absorption would preclude the
detection of these objects at the inferred distance. 

An interpretation as the \hi\ shell of a single SNR is also not tenable.
As  for GSH\,138-01-94 and GSH\,90+03-99, \G\ has likely been caused by
the combined winds and SN explosions of a number of massive stars.

The \hi\ shell appears open to the north, i.e. in a direction away from
the Galactic plane, suggesting that the shell has burst out of the Galactic
disk.  A filament displaced from the northern boundary 
(Filament B, Fig.~\ref{average}) could be the remains of the top (now
missing) part of the shell.
The faint radio continuum emission, consisting
of filaments more or less aligned in a direction perpendicular to the
Galactic plane, lends support to the breakout hypothesis.

Two relatively compact thermal sources (Table~\ref{fluxalpha}), seen
in projection near or on the boundary of the \hi\ shell, could have 
formed in gas compressed by the expanding shell.  One of the DWCL sources,
\IRdos, seen projected inside a smaller \hi\ shell, could also be a product of
triggered star formation.  We cannot rule out the possibility that the
other DWCL sources could also have formed in the same manner.

\begin{acknowledgements}
The research presented in this paper has used data from the Canadian
Galactic Plane Survey, a Canadian project with international partners,
supported by the Natural Sciences and Engineering Research Council.
The work of S.P. and S.C. was supported by the Natural Sciences and
Engineering Research Council of Canada and the Fonds Qu\'ebecois pour
la Recherche sur les sciences de la Nature et la Technologie. 
SP acknowledges the hospitality of the Instituto Argentino de Radioastronom\' \i a
where part of this work was carried out. 
We are grateful to the referee, whose suggestions led to the improvement of this paper.
This project was partially financed by the Consejo Nacional de 
Investigaciones Cient\'{\i}ficas y T\'ecnicas (CONICET) 
of Argentina under projects PIP 01299 and PIP 6433, Agencia PICT 00812 and UBACyT X482.

\end{acknowledgements}

\bibliographystyle{aa}  
\bibliography{bibliografia-1}
   
\IfFileExists{\jobname.bbl}{}
{\typeout{}
\typeout{****************************************************}
\typeout{****************************************************}
\typeout{** Please run "bibtex \jobname" to optain}
\typeout{** the bibliography and then re-run LaTeX}
\typeout{** twice to fix the references!}
\typeout{****************************************************}
\typeout{****************************************************}
\typeout{}
}

\

\end{document}